\documentclass[%
mathleft,%
]{an}
\usepackage{graphicx}
\usepackage{times}
\usepackage{verbatim} 
\usepackage[fleqn]{amsmath}
\usepackage[round,authoryear,sort]{natbib}
\graphicspath{{./plots/}}
\begin{document}

% The following seven commands are intended for editorial usage and
% should be ignored by the author(s).
\Pagespan{1}{}% Document's page range. 
% If second parameter is left empty, the last page is computed
% automatically.
\Yearpublication{2015}%
\Yearsubmission{2015}%
\Month{1}%   
\Volume{999}%  
\Issue{92}% 
% \DOI{This.is/not.aDOI}% 

\title{Tracing the neutral gas environments of young radio AGN with
  ASKAP}

\author{J.\,R.~Allison\inst{1}\fnmsep\thanks{Corresponding author:
    \email{james.allison@csiro.au}} \and E.\,M.~Sadler\inst{2,3} \and
  V.\,A.~Moss\inst{2,3} \and L.~Harvey-Smith\inst{1} \and
  I.~Heywood\inst{1,4} \and B.\,T.~Indermuehle\inst{1} \and
  D.~McConnell\inst{1} \and R.\,J.~Sault\inst{1} \and
  M.\,T.~Whiting\inst{1}}
% Example for footnote, note the usage of the \texttt{fnmsep} command
% as separator between institute number and footnote mark}
\titlerunning{\mbox{H\,{\sc i}} absorption with ASKAP}
\authorrunning{J.\,R. Allison} 
\institute{CSIRO Astronomy \& Space
  Science, P.O. Box 76, Epping, Australia
\and 
Sydney Institute for Astronomy, School of
  Physics A28, University of Sydney, NSW 2006, Australia
\and 
ARC Centre of Excellence for All-sky Astrophysics
  (CAASTRO)
\and
Department of Physics and Electronics, Rhodes University, 
PO Box 94, Grahamstown 6140, South Africa}

\received{XXXX}
\accepted{XXXX}
\publonline{XXXX}


\abstract{At present neutral atomic hydrogen (\mbox{H\,{\sc i}}) gas
  in galaxies at redshifts above $z \sim 0.3$ (the extent of 21\,cm
  emission surveys in individual galaxies) and below $z \sim 1.7$
  (where the Lyman-$\alpha$ line is not observable with ground-based
  telescopes) has remained largely unexplored. The advent of precursor
  telescopes to the Square Kilometre Array will allow us to conduct
  the first systematic radio-selected 21\,cm absorption surveys for
  \mbox{H\,{\sc i}} over these redshifts. While \mbox{H\,{\sc i}}
  absorption is a tracer of the reservoir of cold neutral gas in
  galaxies available for star formation, it can also be used to reveal
  the extreme kinematics associated with jet-driven neutral outflows
  in radio-loud active galactic nuclei. Using the six-antenna Boolardy
  Engineering Test Array of the Australian Square Kilometre Array
  Pathfinder, we have demonstrated that in a single frequency tuning
  we can detect \mbox{H\,{\sc i}} absorption over a broad range of
  redshifts between $z = 0.4$ and $1.0$. As part of our early science
  and commissioning program, we are now carrying out a search for
  absorption towards a sample of the brightest GPS and CSS sources in
  the southern sky. These intrinsically compact sources present us
  with an opportunity to study the circumunuclear region of recently
  re-started radio galaxies, in some cases showing direct evidence of
  mechanical feedback through jet-driven outflows. With the
  sensitivity of the full ASKAP array we will be able to study the
  kinematics of atomic gas in a few thousand radio galaxies, testing
  models of radio jet feedback well beyond the nearby Universe.}

\maketitle

\section{Introduction}

Compact steep spectrum (CSS) and gigahertz-peaked spectrum (GPS) radio
sources have typical sizes much less than the gas distribution in
their host galaxies and so provide us with a means to directly observe
the interaction between radio-jets and the inner region of the
interstellar medium (ISM). By studying the kinematics of the neutral
gas around luminous young radio sources we can test models of early
jet-ISM interaction (e.g. \citealt{Wagner:2012}) that contributes
significantly to AGN feedback in powerful radio galaxies
(e.g. \citealt{Holt:2008, Nesvadba:2008}).

Direct evidence for fast ($> 1000\,\mathrm{km}\,\mathrm{s}^{-1}$) and
massive (up to $50\,\mathrm{M}_{\odot}\,\mathrm{yr}^{-1}$) outflows of
atomic, molecular and ionised gas associated with radio jet-ISM
interaction have now been found in several systems
(e.g. \citealt{Mahony:2013, Morganti:2013,
  Tadhunter:2014}). Absorption from neutral atomic hydrogen
(\mbox{H\,{\sc i}}) gas located in front of the radio source is an
excellent tracer of the kinematics of circumnuclear neutral gas and is
thus an important tool in revealing the presence of neutral gas
outflows (e.g. \citealt{Morganti:2005b, Teng:2013}). Previous 21\,cm
spectroscopic surveys of radio-loud AGN have found that detection
rates of \mbox{H\,{\sc i}} absorption are highest in the most compact
radio galaxies (e.g. \citealt{Morganti:2001, Vermeulen:2003,
  Gupta:2006a, Chandola:2011}) suggesting that these young sources are
cloaked behind high column density gas (\citealt{Pihlstrom:2003,
  Emonts:2010}), but that orientation and geometric effects (including
the size and distribution of the gas clouds) may also play a
significant role (\citealt{Curran:2013b, Orienti:2006}). Comparisons
of the absorption line profiles in larger samples of extended and
compact radio sources show a prevalence of asymmetries and larger
widths ($\Delta{v} \ga 200$\,km\,s$^{-1}$) associated with the latter,
implying that the inner circumunclear medium is significantly
disturbed by younger radio jets clearing their way through the ISM
(e.g. \citealt{Gereb:2014,Gereb:2015}).

\begin{figure*}
\centering
\includegraphics[width = 0.9\textwidth]{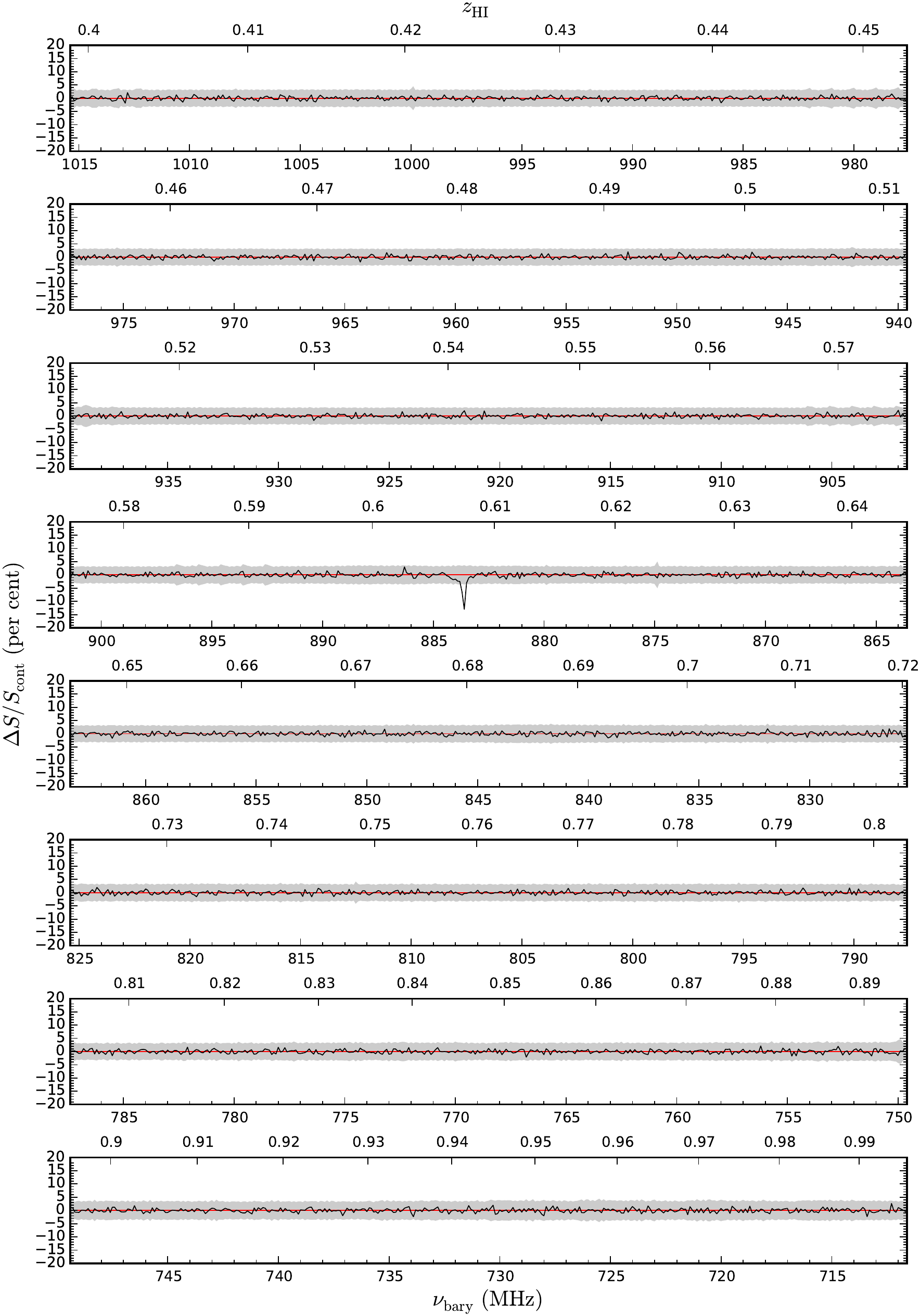}
\caption{Example of an ASKAP-BETA spectrum towards the flat spectrum
  quasar PKS\,B2252$-$089. For visual clarity the data have been
  binned from the native spectral resolution of 18.5\,kHz to
  100\,kHz. The barycentric corrected observed frequency is shown on
  the lower-abscissa, and the upper-abscissa denotes the corresponding
  \mbox{H\,{\sc i}} redshift. The data (black line) denote the
  absorbed fraction of background continuum and the grey region gives
  the corresponding RMS spectral noise multiplied by a factor of
  5. The absorption line, first detected by \cite{Curran:2011a} using
  the Green Bank Telescope, is visible in the spectrum at
  $\nu_\mathrm{bary} = 883.6$\,MHz, equal to an \mbox{H\,{\sc i}}
  redshift of $z = 0.6076$.}\label{figure:PKS2252-089_full_spectrum}
\end{figure*}

Within the next decade, the advent of the Square Kilometre Array (SKA)
pathfinder telescopes will enable astronomers to carry out
radio-selected \mbox{H\,{\sc i}} absorption surveys towards a few
hundred thousand sources over the entire sky out to redshifts greater
than $z \sim 1$ (\citealt{Morganti:2015}). Larger samples of distant
sources can be constructed and compared with more well studied nearby
systems (e.g. \citealt{Maccagni:2014}), greatly aiding our
understanding of the typical gaseous environments of distant GPS and
CSS sources and their evolution with redshift. Further interpretation
of the detected systems will be supported by follow-up observations
with millimetre (e.g. the Atacama Large Millimetre Array) and optical
facilities (e.g. the Very Large Telescope), providing confirmation of
association through spectroscopic redshifts. Furthermore, 21\,cm
spectroscopy at higher spatial resolution using Very Long Baseline
Interferometry (VLBI) will allow individual modelling of the
distribution and kinematics of the atomic gas. Here we discuss some of
the first results from commissioning of the Australian SKA Pathfinder
(ASKAP; \citealt{Deboer:2009, Schinckel:2012}), demonstrating the wide
fractional bandwidth and exceptionally radio quiet site of this
telescope.

\section{\mbox{H\,{\sc i}} absorption with ASKAP}

ASKAP will be a 36-antenna radio interferometer operating at
frequencies between 700 and 1800\,MHz. Each antenna is equipped with a
Phased Array Feed (PAF) ``chequerboard'' array receiver
(\citealt{Hay:2008}), containing 188 independent receptors that can be
used to form up to 36 primary beams over a 30 square degree
field-of-view. This will allow astronomers to perform rapid surveying
of the cm-wavelength sky. Among several surveys of the entire southern
sky, the ASKAP First Large Absorption Survey in \mbox{H\,{\sc i}}
(FLASH) will search for cool atomic hydrogen towards 150\,000 southern
($\delta \la +10$\degr) radio sources, finding a few thousand
intervening and associated systems out to $z = 1.0$.

The telescope is currently in its science demonstration and
commissioning phase comprising 6 antennas fitted with the first
generation of PAFs, forming the Boolardy Engineering Test Array (BETA;
\citealt{Hotan:2014}). In its lowest frequency band between 711.5 and
1015.5\,MHz, the BETA telescope enables us to search for \mbox{H\,{\sc
    i}} absorption at redshifts between $z = 0.4$ and 1.0 against the
brightest ($S_{1.4} \sim 1$\,Jy) radio sources in the southern
sky. The 16\,416 spectral channels, separated by 18.5\,kHz, give a
velocity resolution between 5.5 and 7.8\,km\,s$^{-1}$ over this
band. With 6 antennas the array has a 5\,$\sigma$ sensitivity per 2\,h
on-source (at 1\,GHz) of about 10\,per\,cent absorption against
sources brighter than 1\,Jy. Assuming a cold-phase 21\,cm spin
temperature of 100\,K and line width of 30\,km\,s$^{-1}$ this equates
to an \mbox{H\,{\sc i}} column density of approximately
$5\times10^{20}$\,cm$^{-2}$. With the full ASKAP complement of 36
antennas this sensitivity will increase to sources brighter than
150\,mJy, greatly increasing the number of available targets. Even
with its lower sensitivity, BETA still provides us with a unique
opportunity to carry out new science over a largely unexplored epoch.

\begin{figure}
\centering
\includegraphics[width = 0.45\textwidth]{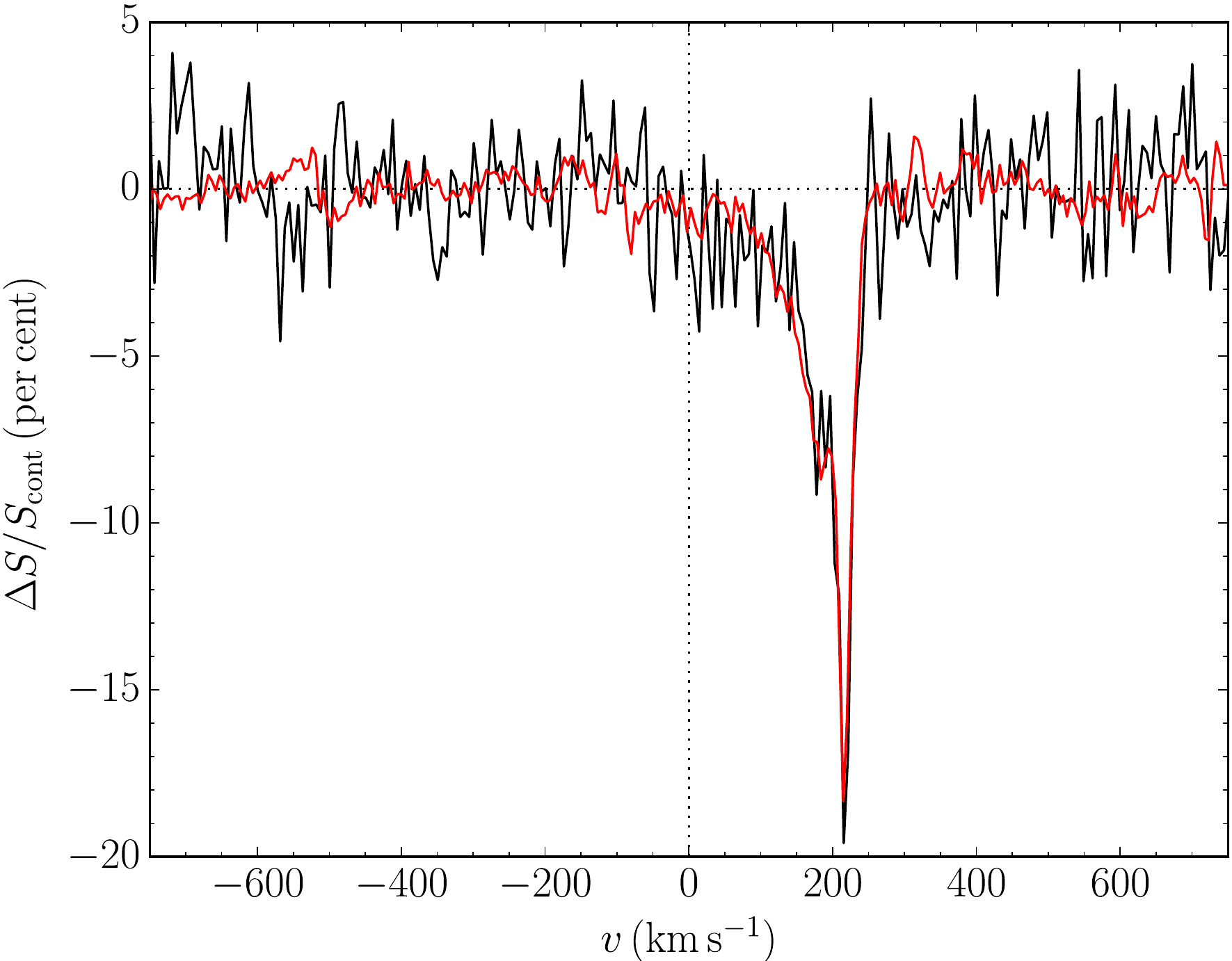}
\caption{Comparison of the BETA spectrum (black) and GBT spectrum
  (red; \citealt{Curran:2011a}) towards PKS\,B2252$-$089. The velocity
  is given with reference to the optical spectroscopic redshift of $z
  = 0.6064$ (\citealt{Drinkwater:1997}). The BETA spectrum reproduces
  the broad feature seen bluewards of the peak
  absorption.}\label{figure:PKS2252-089_comparison_spectrum}
\end{figure}

In Fig.\,\ref{figure:PKS2252-089_full_spectrum} we show an example
spectrum towards the flat spectrum quasar PKS\,B2252-089 from an
8.5\,h on-source integration with five of the BETA antennas. The
source has a continuum flux density of 1\,Jy at 800\,MHz. This example
spectrum demonstrates the power of the telescope for detecting
\mbox{H\,{\sc i}} gas in absorption, with no strong radio frequency
interference evident in this band. The absorption feature originally
detected with the Green Bank Telescope (\citealt{Curran:2011a}) and
associated with the host galaxy of the radio source is clearly evident
at a frequency of 883.6\,MHz. The striking agreement between the BETA
and GBT spectra can be seen in
Fig.\,\ref{figure:PKS2252-089_comparison_spectrum}, with the former
spectrum confirming the broad wing seen bluewards of the peak
absorption. The \mbox{H\,{\sc i}} is redshifted by 200\,km\,s$^{-1}$
with respect to the optical [O\,III] λλ4959, 5007 emission lines
(\citealt{Drinkwater:1997}), which might be indicative of an inflow of
cool neutral gas towards the nucleus along the
line-of-sight. Alternatively if the peak absorption against the
compact radio source is a more accurate indicator of the AGN redshift,
then the blue wing would suggest that the radio jets in this source
(\citealt{Liu:2002} and references therein) could be driving the
atomic gas in an outflow.

\section{A pilot 21\,cm survey of GPS/CSS sources at cosmological
  redshifts}

The sub-kpc sizes of GPS and CSS sources make them ideal targets for a
pilot \mbox{H\,{\sc i}} absorption survey with BETA, where the
continuum radio emission is well matched to the expected extent of
foreground gas clouds. Given the limited sensitivity of the 6-antenna
BETA telescope, we are carrying out observations of bright ($S >
1.5$\,Jy) southern GPS/CSS sources selected from the unbiased sample
of \cite{Randall:2011}. These sources represent some of the most
powerful young or re-started radio galaxies in the Universe. Of the 26
Randall et al. sources we selected 13 with expected \mbox{H\,{\sc i}}
redshifts in the 711.5 -- 1015.5\,MHz band, i.e. redshifts between 0.4
and 1.0. Some of these redshifts are only indicative values from
optical photometry, but the large fractional bandwidth available with
BETA allows us to blindly search for \mbox{H\,{\sc i}} and OH
absorption over a wide range of redshifts.

\begin{figure}
\centering
\includegraphics[width = 0.45\textwidth]{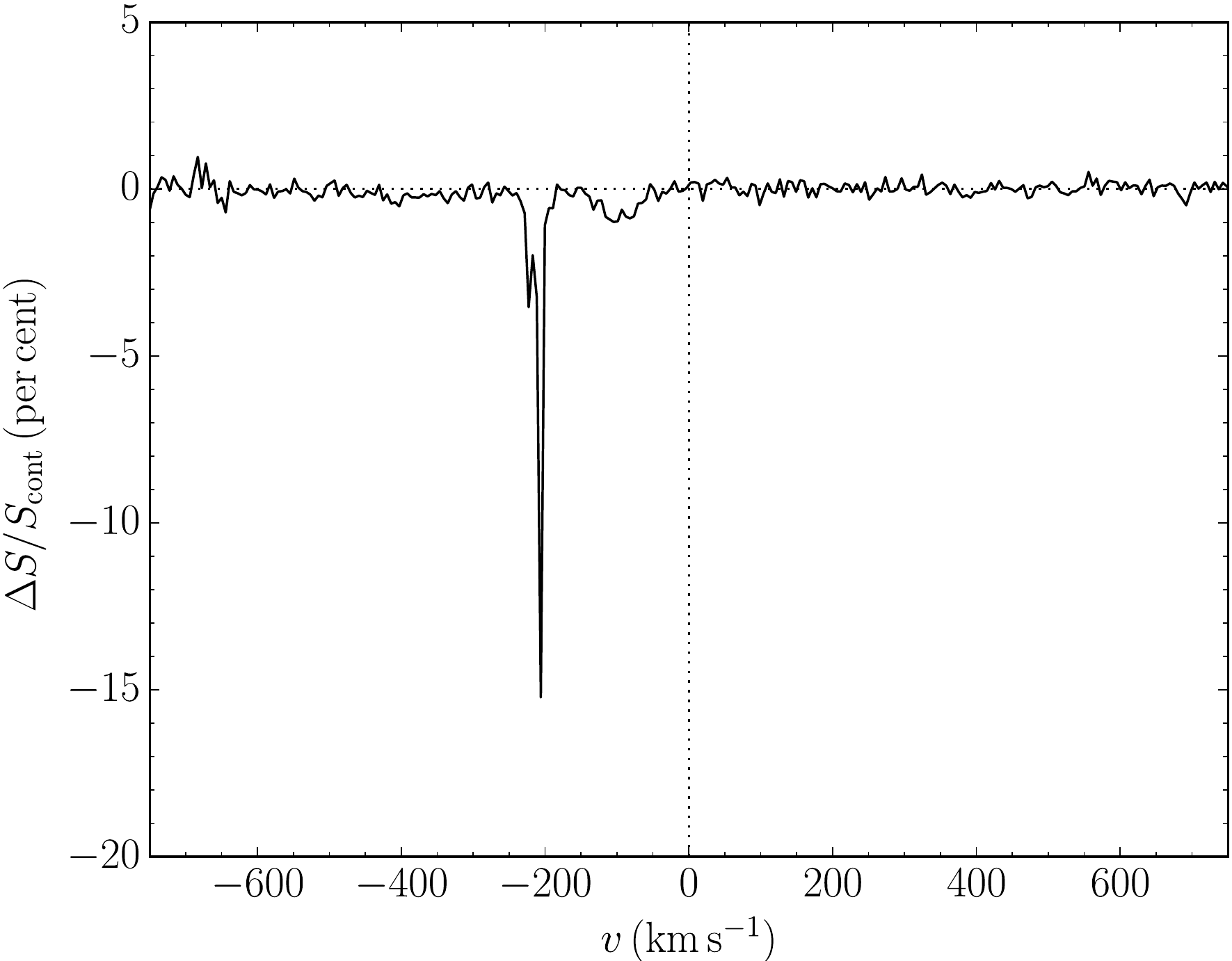}
\caption{Recently discovered \mbox{H\,{\sc i}} absorption in the host
  galaxy of the powerful GPS radio source PKS\,B1740$-$517. The
  velocity is given with reference to the optical redshift of $z =
  0.44230\pm0.00022$
  (\citealt{Allison:2015}).}\label{figure:PKS1740-517_spectrum}
\end{figure}

At present our observations with the BETA telescope are ongoing, but
we have recently achieved our first discovery of \mbox{H\,{\sc i}}
absorption with the ASKAP-BETA telescope (\citealt{Allison:2015}). The
background radio source, PKS\,B1740$-$517, is in the
\cite{Randall:2011} sample and has a gigahertz-peaked spectrum with a
spectral age of approximately 2500\,yrs. The absorption seen at $z =
0.4413$ is characteristically complex, with a deep narrow component
that is almost unresolved at 5\,km\,s$^{-1}$ and broader components
with more typical widths (see
Fig.\,\ref{figure:PKS1740-517_spectrum}). The complexity of this
profile is indicative of gas associated with the host galaxy, where
the different velocity components trace either the spatial structure
of the radio source, or the gas kinematics caused by the
AGN. Association with the host galaxy was confirmed by optical
spectroscopy from follow up observations with the Gemini South
Telescope, providing the first spectroscopic redshift for this
source. This young radio source is likely shrouded within a dense
gaseous environment through which we are seeing absorption of the
continuum, an interpretation that is supported by the strong
absorption of soft X-ray emission seen in archival \emph{XMM-Newton}
data.

\section{Summary}

Absorption of continuum radio emission by cool \mbox{H\,{\sc i}} gas
located in the foreground allows us to directly observe the
interaction between radio AGN and neutral gas in the surrounding
ISM. With \mbox{H\,{\sc i}} absorption seen in the spectra of the most
compact radio sources we can probe further into the circumnuclear
medium and test models of jet-driven feedback in the early stages of
radio-jet growth. With our first commissioning observations using the
ASKAP BETA telescope we are searching for \mbox{H\,{\sc i}} absorption
over a continuous redshift range between $z = 0.4$ and $1.0$, a range
that was previously unobtainable due to smaller available fractional
bandwidths and poorer radio frequency environments. We are now
carrying out a pilot survey of the brightest and most compact radio
sources in the southern hemisphere, including redshifted GPS/CSS
sources selected from the sample of \cite{Randall:2011}. Recently we
made our first discovery of \mbox{H\,{\sc i}} absorption towards the
GPS radio source PKS\,B1740$-$517 (\citealt{Allison:2015}), revealing
the dense neutral gas surrounding this young radio galaxy. With the
full capability of ASKAP we will be able to carry out FLASH (the First
Large Absorption Survey in \mbox{H\,{\sc i}}) towards 150\,000 radio
sources across the entire southern sky. This will provide over a
thousand detections of associated absorption and allow us to explore
the evolution of jet-ISM feedback in young radio AGN as a function of
redshift.

\acknowledgements

We thank the anonymous referee for useful comments that helped improve
this paper.

The Australian SKA Pathfinder is part of the Australia Telescope
National Facility which is managed by CSIRO. Operation of ASKAP is
funded by the Australian Government with support from the National
Collaborative Research Infrastructure Strategy. Establishment of the
Murchison Radio-astronomy Observatory was funded by the Australian
Government and the Government of Western Australia. ASKAP uses
advanced supercomputing resources at the Pawsey Supercomputing
Centre. We acknowledge the Wajarri Yamatji people as the traditional
owners of the Observatory site.

JRA acknowledges support from a Bolton Fellowship. This work was
supported by iVEC, through the use of advanced computing resources
located at the Pawsey Centre. We have made use of \texttt{Astropy}, a
community-developed core \texttt{Python} package for Astronomy
(Astropy Collaboration, 2013); the NASA/IPAC Extragalactic Database
(NED) which is operated by the Jet Propulsion Laboratory, California
Institute of Technology, under contract with the National Aeronautics
and Space Administration; NASA's Astrophysics Data System
Bibliographic Services; the SIMBAD data base and VizieR catalogue
access tool, both operated at CDS, Strasbourg, France.

\end{document}